\normalfont
\documentclass[%
 aip,
 amsmath,amssymb,
 reprint,
groupedaddress,%
]{revtex4-1}

\usepackage{graphicx}
\usepackage{dcolumn}
\usepackage{bm}
\usepackage{verbatim} 
\usepackage{gensymb} 
\usepackage{epstopdf} 

\begin{document}

\preprint{AIP/123-QED}

\title{In-Vacuum Active Electronics for Microfabricated Ion Traps}

\author{Nicholas D. Guise}
\email{nicholas.guise@gtri.gatech.edu}
\author{Spencer D. Fallek}
\author{Harley Hayden}
\author{C-S Pai}
\author{Curtis Volin}
\author{K. R. Brown}
\author{J. True Merrill}
\author{Alexa W. Harter}
\author{Jason M. Amini}
\affiliation{Georgia Tech Research Institute,  Atlanta, Georgia 30332, USA}

\author{Lisa M. Lust}
\author{Kelly Muldoon}
\author{Doug Carlson}
\author{Jerry Budach}
\affiliation{Honeywell International, Plymouth, Minnesota 55441, USA}

\date{\today}

\begin{abstract}
The advent of microfabricated ion traps for the quantum information community has allowed research groups to build traps that incorporate an unprecedented number of trapping zones. However, as device complexity has grown, the number of digital-to-analog converter (DAC) channels needed to control these devices has grown as well, with some of the largest trap assemblies now requiring nearly one hundred DAC channels.  Providing electrical connections for these channels into a vacuum chamber can be bulky and difficult to scale beyond the current numbers of trap electrodes. This paper reports on the development and testing of an in-vacuum DAC system that uses only 9 vacuum feedthrough connections to control a 78-electrode microfabricated ion trap.  The system is characterized by trapping single and multiple $^{40}\text{Ca}^+$ ions.  The measured axial mode stability, ion heating rates, and transport fidelities for a trapped ion are comparable to systems with external (air-side) commercial DACs.
\end{abstract}

\maketitle

\section{\label{sec:intro}Introduction}

Trapped ions are one of the most promising platforms for implementing atomic control and quantum information processing. Varying quasi-static potentials on trap electrodes can transport the ions while preserving information stored in the atomic state. Laser fields and oscillatory magnetic fields can manipulate the atomic states of single ions or couple the information between two or more ions.  In the transport architecture for a quantum information processor \cite{Kielpinski2002}, large numbers of ions, each representing one quantum bit of information, can be transported between many interaction zones where pairs of ions can interact.

The advent of microfabricated ion traps and, in particular, surface-electrode structures that can be fabricated on single substrates \cite{Chiaverini2005,Seidelin2006} allows the complexity of ion traps to grow to unprecedented levels.  Larger traps require nearly one hundred applied potentials, each typically supplied by a digital-to-analog converter (DAC), to fully control the positions of the ions \cite{Wright2013}. However, providing the wiring for these complex traps is unwieldy, as traditional through-vacuum feedthroughs are bulky and can introduce noise via stray inductance and capacitance. Recent efforts have reduced the wiring complexity using vacuum chambers that incorporate the ceramic trap carriers as part of the vacuum housing or use in-vacuum circuit boards to distribute and filter signals \cite{Doret2012,Graham2013,Allcock2012,Blakestad2011,Moehring2011}. Both techniques simplify the wiring of the DAC signals, but they do not reduce the total number of vacuum-to-air connections needed to supply these signals.

A novel alternative presented here is to place the DACs inside the vacuum system, replacing bundles of analog wires with a few digital serial lines that can address the individual DAC channels.  Integration with surface-electrode ion traps places strict requirements on such an in-vacuum DAC system. The DACs must not introduce excess noise on the ion trap electrodes, since this can cause ion heating. Voltage updates must be sufficiently fast and smooth for ion transport operations. Physical layout of the in-vacuum DACs and their support components (power, filters, etc.) must not obstruct laser access to the surface trap circuit board, as required for ion manipulation.  Finally, the in-vacuum DAC materials, including the die packaging, circuit boards, and component solders, must meet the ultra-high vacuum (UHV) requirements for ion trapping ($10^{-11}$ torr or lower).

Here we describe an integrated ion trapping system (Fig. \ref{fig:Overview}) consisting of a 78-electrode trap with direct current (DC) potentials provided by two 40-channel in-vacuum DAC chips. The DACs are programmed via a four line serial bus (1 clock, 1 bit align, and 2 serial data lines). The DACs are decapsulated commercial components, packaged together with the ion trap into a compact assembly with two three-layer circuit boards. The assembly installs by insertion into an in-vacuum edge connector socket built into the vacuum chamber.  We demonstrate ion trapping and transport and measure the ion heating rate and axial mode stability.

This paper is organized as follows. Design, specifications, and fabrication of the in-vacuum electronics are described in Sec. \ref{sec:electronics}. Control software and timing considerations are presented in Sec. \ref{sec:software}.  Integration with the microfabricated ion trap is described in Sec. \ref{sec:integration}.  Section \ref{sec:testing} presents results from testing and characterization of the integrated system, with performance comparisons to standard air-side DACs.

\section{\label{sec:electronics}In-Vacuum Electronics}

\subsection{\label{sec:config}Board Layout}

\begin{figure}[htbp]
\includegraphics{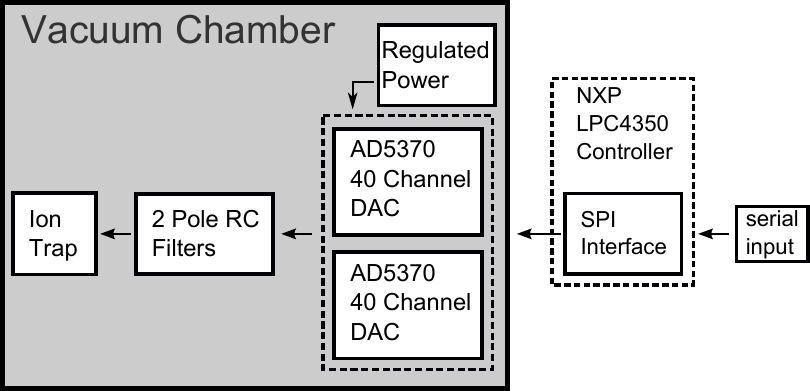}
\caption{\label{fig:Overview}Architecture overview for the integrated system.}
\end{figure}

Considerations for the in-vacuum control electronics include update speed, voltage noise, number of electrical feedthroughs, and flexibility to address channels individually rather than through a global update. The chosen architecture (Fig. \ref{fig:Overview}) locates two 40 channel DACs and a set of low-pass RC filters near the ion trap in the vacuum system. The selected DACs, Analog Devices \footnote{Mention of specific devices or manufacturers herein does not constitute implied or explicit endorsement by GTRI or Honeywell International.} AD5730's (40-channel, 16-bit), are capable of $\pm10$ V operation using a +5 VDC stable reference and $\pm12$ VDC supplies. We measure an unfiltered spectral noise density of 15 nV$/\sqrt{\text{Hz}}$ at 1 MHz. The architecture reduces the required number of UHV feedthroughs to 9 essential control lines plus a separate feedthrough for radio frequency (RF) power. The design scales well to larger numbers of electrodes; each additional serial data line controls an additional 40 DAC channels.

\begin{figure}[htbp]
\includegraphics{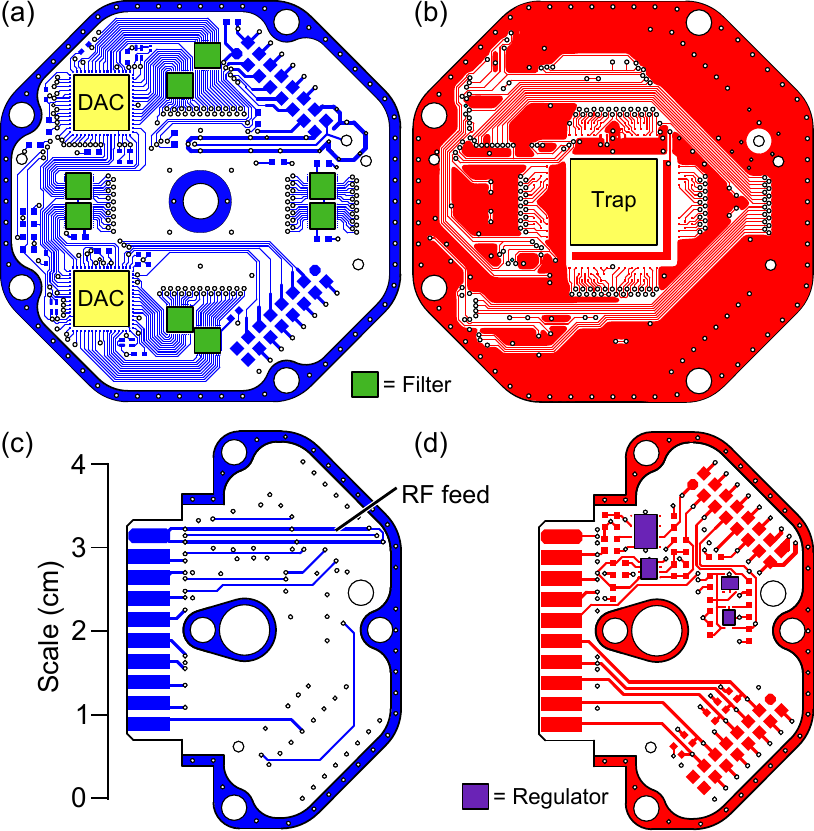}
\caption{\label{fig:IEMITlayout}Board layouts: (a) trap board bottom; (b) trap board top; (c) regulator board bottom; (d) regulator board top.}
\end{figure}

The in-vacuum circuitry is configured in a dual board layout. The trap board (Fig. \ref{fig:IEMITlayout}a-b) contains the microfabricated ion trap as the sole component on its top  surface, with the DAC chips and RC filters on its bottom surface. This arrangement keeps the critical active electronics on the backside of the board, avoiding direct exposure to laser light which could modify semiconductor components due to carrier generation or cause unwanted photon scattering during ion trap operations.  The trap RF line is a 100 $\Omega$ grounded coplanar transmission line, routed to minimize radiative crosstalk.

The regulator board (Fig. \ref{fig:IEMITlayout}c-d) takes $\pm14$ VDC and generates the $\pm12$ VDC DAC supply voltages, a +3.3 VDC logic supply, and a +5 VDC ultra-stable reference. All connections to in-vacuum cabling are made through a $2 \times 10$ pin card edge connector on the regulator board; placement of this connector on the trap board would over-restrict laser access to the trap surface.  The trap RF, bipolar power supply, ground, and serial lines are supplied through the edge connector. Digital and analog sections are segregated to minimize cross-coupling. The digital lines are supplied as twisted pairs terminated with 100 $\Omega$ resistors on the regulator board.  Decoupling 1 nF capacitors are added to several supply lines to further suppress RF pickup.  The regulator board is connected to the trap board with spring-loaded pins and polyether-ether-ketone (PEEK) board stackers (see Sec. \ref{sec:integration}).

\subsection{\label{sec:components}Board Components}
The trap and regulator circuit board material is a multilayer Rogers 4350B patterned substrate. This material is a UHV-compatible circuit board that has been used successfully for mounting operational ion traps \cite{Shappert2013,Tanaka2009}.  The multilayer structure is constructed with 4450 prepreg, an uncured form of the 4350B board.  Eight plies of 4450 prepreg are used to form a 3-metal-layer stack with a single internal ground plane.  The multilayer construction adds stiffness and provides an internal ground plane to reduce crosstalk between the top and bottom layers. To reduce outgassing, no soldermask is applied. The boards are 0.065'' thick with soft bondable gold finish over 1 oz copper.  The entire soldering process is kept at or below the 4350B glass transition temperature of 280\celsius\hspace{2 pt}.

The AD5370's and regulators are decapsulated from the packaged component to bare die; this is required to prevent outgassing of the molding plastic in the UHV ion trap environment. A jet etching technique with fuming nitric acid (HNO$_3$) and sulfuric acid (H$_2$SO$_4$) dissolves the encapsulant on the package, leaving the die, bond pads, and wire bonds intact. Details of a similar decapsulation process are available in Ref. \citenum{Murali2006}.

The RC low-pass components are second-order filters, obtained from Semiconwell as a custom order of their thin-film tapped EMI/RFI filter product line. Each 0.10''$\times$0.12'' die contains 12 two-pole RC filters. The resistors are 35 k$\Omega$ thin film tantalum-nitride and the capacitors are 220 pF metal-nitride-oxide-silicon.  The two RC stages are identical; filter performance could in principle be improved by choosing different impedance values to minimize loading of the second stage by the first. 

During probing and wirebonding, delicate handling of the filter pads is required to avoid damage to the thin silicon nitride beneath the pad.  The filter function is found to depend on the bias voltage in certain regimes, due to a metal-oxide-semiconductor capacitance effect caused by insufficient doping of the substrate. A constant RC response ($f_{\text{3dB}}=12.1$ kHz) is obtained by back-biasing the substrate at $-12$ V, leaving the device in the accumulation state over the full range of applied bias voltages $-10$ V to $+10$ V.  In this regime the overall filter capacitance is given by $C_{\text{oxide}}=220$ pF. 

\subsection{\label{sec:construction}Construction}
\begin{figure}[htbp]
\includegraphics{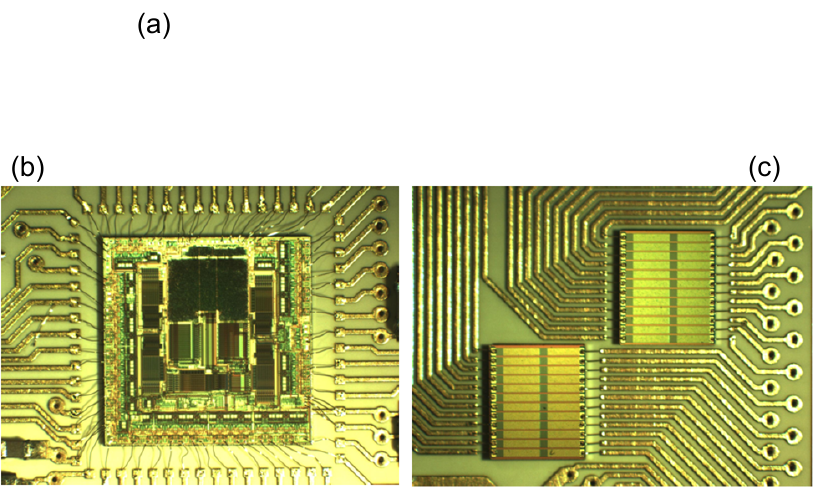}
\caption{\label{fig:BondingJoints}(a) 0402 component soldered with a flux free process; (b) decapsulated DAC die bonded with a thermocompression process; (c) filter dies bonded with a gold ball process.}
\end{figure}

The trap and regulator boards are ultrasonically cleaned (in acetone, isopropyl alcohol, and deionized water) prior to soldering components. Passive capacitors and resistors are pre-tinned with SAC305 eutectic (96.5\% Sn, 0.5\% Cu, 3\% Ag) and hand soldered to the boards using a gold-plated soldering needle. To minimize the use of high-outgassing materials, no flux or soldermask are used.  The soldering process is facilitated by placing the board on a 170\celsius\hspace{2 pt} hot stage, allowing the use of an ultra fine needle tip which otherwise cannot transfer sufficient heat to the solder (eutectic point 217\celsius). Each connection is inspected for brittleness; Fig. \ref{fig:BondingJoints}a shows an example solder joint. Once fully soldered, the boards are again ultrasonically cleaned to remove tin oxide particulate.

\begin{figure}[htbp]
\includegraphics{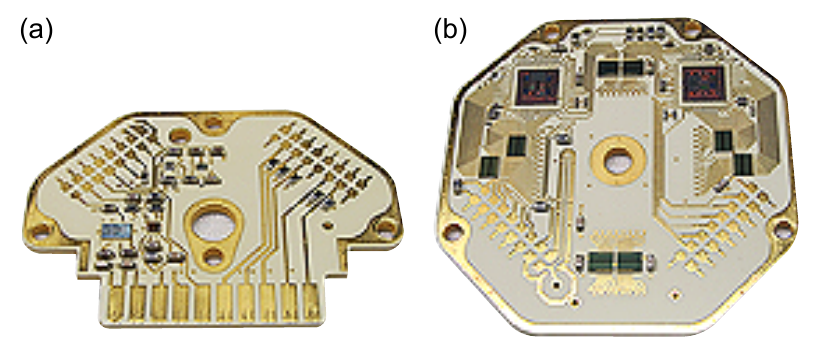}
\caption{\label{fig:BoardPhotos}(a) Regulator and (b) trap boards with all passives and decapsulated ICs attached.}
\end{figure}

On the regulator board, all integrated circuit (IC) die are adhered with low outgassing silver-filled epoxy (EPO-TEK H21D). The epoxy is cured at 120\celsius\hspace{2 pt} for 15 minutes.  On the trap board the two DAC die and 8 filter banks are adhered with the H21D epoxy.  The decapsulated DAC die are wirebonded using a thermocompression bonder (Fig. \ref{fig:BondingJoints}b); gold ball bonding is used to wirebond the filter die (Fig. \ref{fig:BondingJoints}c).  With all passives and ICs connected, the boards are cleaned in isopropyl alcohol and deionized water rinse.  The microfabricated ion trap is adhered with H21D epoxy, and its electrodes are wirebonded to the trap board electronics.  Figure \ref{fig:BoardPhotos} shows photos of the completed trap and regulator boards. Section \ref{sec:integration} describes assembly of the boards into in-vacuum housing.

\section{\label{sec:software}Control Software}

\begin{figure}[htbp]
\centering
\includegraphics{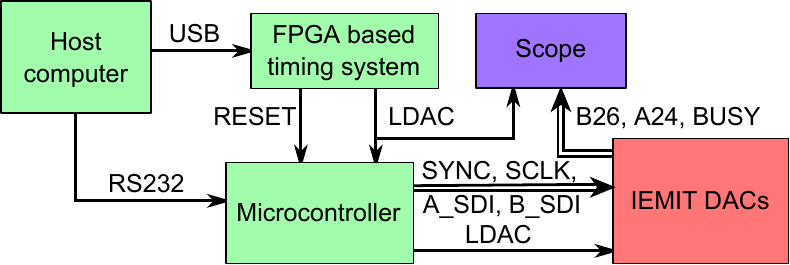}
\caption{\label{fig:IEMITIntegration}Integrated data acquisition system.}
\end{figure}

For ion trap control, communication functions with the in-vacuum DACs are integrated into a data acquisition system as shown in Fig. \ref{fig:IEMITIntegration}. The microcontroller (NXP 204MHz LPC4350) programs the in-vacuum DACs via a serial bus.  Nine signal and power lines run from the controller box through vacuum feedthroughs to the in-vacuum electronics: serial-digital interface (SDI) programming lines for each DAC (A\_SDI and B\_SDI), serial clock (SCLK), bit align (SYNC), load data (LDAC), DAC busy (BUSY), $\pm14$ VDC supply lines, and GND.  The controller utilizes general purpose input/output (GPIO) lines on the microcontroller instead of the built-in serial peripheral interface (SPI) lines, enabling a single common clock to be shared between the two 40 channel DAC chips.

During operation, the DAC system pulls approximately 33 mA from each of the two supply lines for a total of 1 W dissipation. The system pulls an additional 8 mA from the +14 V supply during DAC updates, adding 0.1 W during ion transport. The exterior housing temperature increases to 47\celsius\hspace{2 pt} during normal operation in vacuum. 

\begin{figure}[htbp]
\centering
\includegraphics{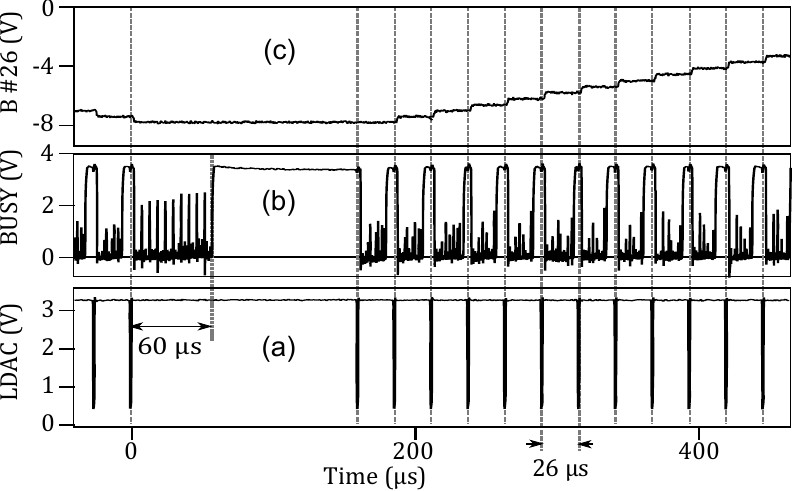}
\caption{\label{fig:TransportTiming}Typical data transfer to in-vacuum DACs for ion transport: (a) sequence of LDACs sent by the timing system; (b) resulting DAC BUSY signals (low when receiving data); (c) output from DAC B26, a diagnostic channel not connected to any trap electrode.}
\end{figure}

The controller receives data via RS232 at 115200 baud.  Each waveform consists of a series of ``packets'' to be sent to the in-vacuum DACs; each packet contains a list of DAC channels and voltages to update. The first packet is sent to the DACs once the data for the entire waveform is loaded into the microcontroller.  Externally supplied LDAC pulses from the field-programmable gate array (FPGA) timing system are used to trigger subsequent voltage updates on the DAC chips  (Fig. \ref{fig:TransportTiming}). Each LDAC pulse triggers the controller to send the next packet to the DACs after a short delay to allow the DACs to latch the previous packet data. A shared BUSY line from both DACs indicates their status (idle or receiving data). 
 
To speed up the update rate for the DACs, data transfer is divided into an initial packet that covers all 40 channel pairs (DACs A and B are programmed in parallel) followed by smaller packets that only cover those channels that have changed since the last update. It takes roughly 60 $\mu$s for the controller to upload 40 channel pairs to the DACs, so the timing system must provide a 60 $\mu$s delay before sending the LDAC that triggers the DACs to use this initial packet. For ion transport, subsequent packets need to change only 8 channel pairs and thus require only 25 $\mu$s to upload. In combination with the 1 $\mu$s LDAC pulse, this corresponds to a 38 kHz update rate. 

\section{\label{sec:integration}Ion Trap Hardware and Integration}

\begin{figure*}[htbp]
\includegraphics{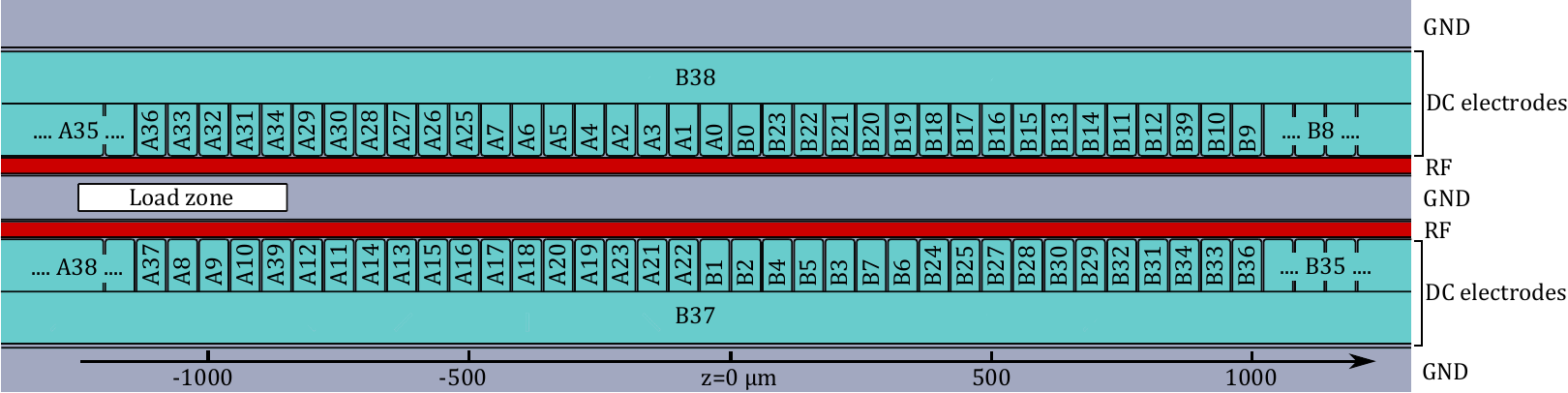}
\caption{\label{fig:GenVTrap}Schematic of the GTRI Gen V trap, showing RF electrodes (red), ion loading slot (white), ground planes (gray), and DC electrodes (teal).  Axially segmented electrodes A0-B36 and radial electrodes B37-B38 are labeled by their control DAC chip (A or B) and channel (0-39).}
\end{figure*}

A new linear ion trap, the Georgia Tech Research Institute (GTRI) Gen V (Fig. \ref{fig:GenVTrap}), was fabricated to demonstrate the in-vacuum DAC system.  Fabrication processes for the Gen V trap are similar to those described for the GTRI Gen II trap in Ref. \citenum{Doret2012}. The trap as designed contains 86 DC electrodes; for this work a number of non-critical electrodes on the ends of the trap are shorted together so that only 78 DAC channels are required.  These are supplied entirely by the two in-vacuum AD5730 DACs, with two channels reserved for diagnostics.

\begin{figure}[htbp]
\centering
\includegraphics{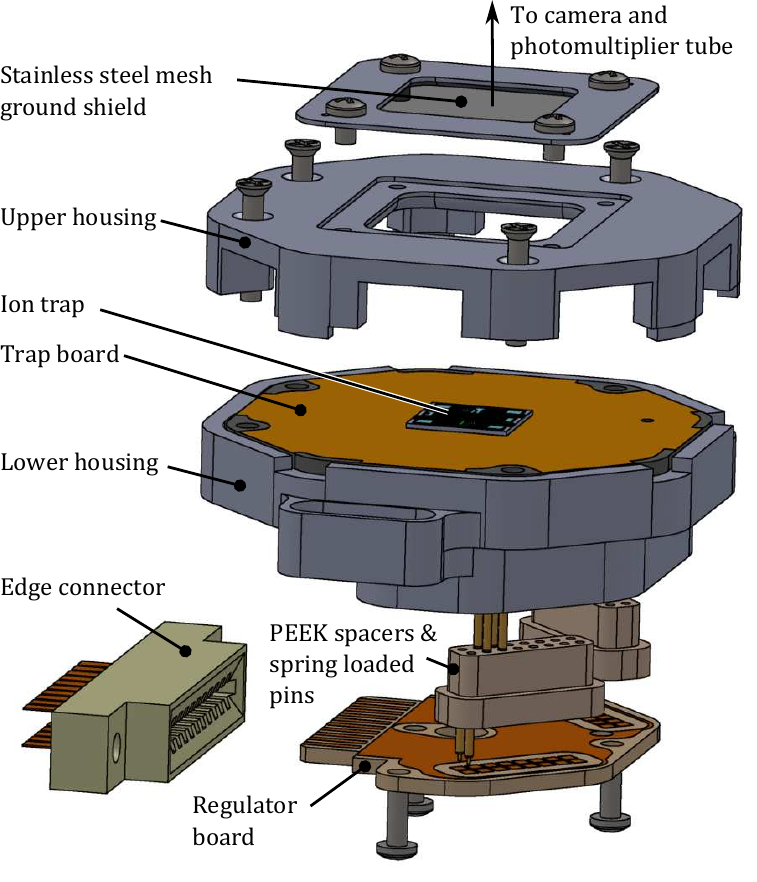}
\caption{\label{fig:IEMITAssembly} Assembly view of housing with PEEK edge connector.}
\end{figure}

The trap housing is machined from 316 stainless steel. Figure \ref{fig:IEMITAssembly} shows an exploded view of the in-vacuum package. The two-board design (see Sec. \ref{sec:config}) introduces the additional complexity of providing interconnects between the trap and regulator boards. Spring-loaded contact pins (Interconnect Devices Inc., part 100785-002) are chosen for their vacuum compatible materials, an overall length that allows placement of the regulator board (with attached edge connector) well below the trap board, and a high pin density that permits multiple ground connections between the boards. The pins are supported and spaced by two PEEK plastic spacers that are glued with H21D epoxy into rectangular slots in the housing.

The ion trap package is designed to fit into a 4.5'' spherical octagon (Kimball Physics MCF450-SphOct-E2A8). The PEEK edge connector (Sullins WMC10DTEH) is mounted to grooves in the octagon using custom “groove grabbers” (Kimball Physics). The groove grabbers also support a frame to hold the neutral oven and a shield to protect the trap from overspray from the oven. The chamber is pumped by a 20 L/s ion pump and a titanium sublimation pump.

The in-vacuum edge connector is a $2\times10$ configuration with ten contacts for each side of the inserted regulator board. The contacts on the edge connector are soldered to Kapton-coated UHV vacuum wire with AuSn eutectic solder. Each signal wire is bundled with a ground wire into a twisted pair. This provides a roughly 100 $\Omega$ characteristic impedance that is matched by resistive termination on the regulator board. The other end of the wiring is connected to a 25-pin PEEK D-sub connector that mates with a vacuum feedthrough.  

Prior to installation of the trap and regulator boards, the vacuum chamber along with PEEK spacers and PEEK edge connector are prebaked in vacuum at 240\celsius\hspace{2 pt} for 470 hours. After installation in the vacuum chamber, the package is baked in vacuum at 200\celsius\hspace{2 pt} for 50 hours.

\section{\label{sec:testing}Testing and Characterization}

\subsection{\label{sec:verification}DAC communication verification}

\begin{figure}[htbp]
\centering
\includegraphics{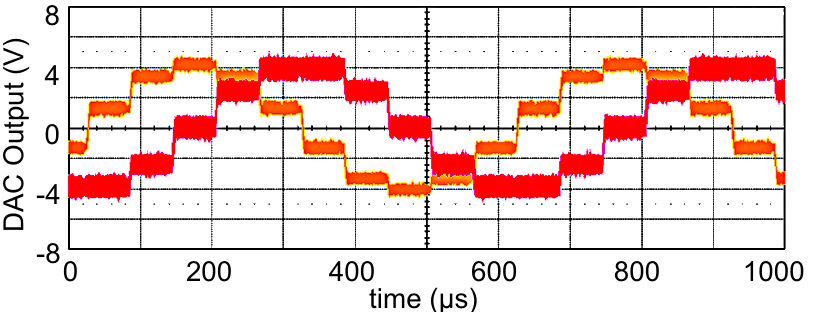}
\caption{\label{fig:EnvelopeTest} $10^4$ traces of a discretized sine and cosine signal generated by the in-vacuum DACs. The noise on the traces is due mostly to RF pickup on the diagnostic lines at the edge connector.}
\end{figure}

RF crosstalk may introduce bit errors in communication with the in-vacuum DACs.  We place bounds on this effect by observing two diagnostic DAC channels not connected to any electrode.  A voltage waveform is sent repeatedly to the diagnostic channels and the resulting DAC potentials are recorded on a digital oscilloscope  (Fig. \ref{fig:EnvelopeTest}). After $10^4$ recorded traces while applying RF (300 Vpp, 37 MHz) to the trap, no measurable update error is observed.  This corresponds to a bit error rate of $2\times10^{-7}$ or lower. The RF noise pickup observed in Fig. \ref{fig:EnvelopeTest} is not expected to appear on DAC lines going to the trap.

\subsection{\label{sec:loading}Ion loading}

$^{40}\text{Ca}^+$ ions are loaded and confined in the trap using standard techniques described in greater detail elsewhere \cite{Haffner2008,Doret2012,Vittorini2013}.  Neutral Ca atoms are supplied by an oven located below the trap board.  The atoms enter through the loading slot (Fig. \ref{fig:GenVTrap}) and are photoionized 60 $\mu$m above the trap surface. Ions are confined radially via a ponderomotive pseudopotential generated by applying RF to the RF electrodes. Potentials applied to the DC electrodes by the DACs confine the ion axially at the desired location along the trap.  Ions are transported by varying the DC trapping potentials to move the location of the potential minimum. Trapped ions are detected and cooled with the ${^2S_{1/2}} \rightarrow {^2P_{1/2}}$ cycling transition at 397 nm. An additional beam at 866 nm repumps ions from the metastable ${^2D_{3/2}}$ level. All laser beams are positioned parallel to the plane of the ion trap. Ion fluorescence emitted perpendicular to the trap plane passes through the mesh shield (Fig. \ref{fig:IEMITAssembly}) and is focused by a lens assembly onto a photomultiplier tube and a charge-coupled device (CCD) camera. The $^{40}\text{Ca}^+$ storage lifetime with Doppler cooling is as long as several hours.

\subsection{\label{sec:transport}Ion dark lifetime and transport}

\begin{figure}[htbp]
\centering
\includegraphics{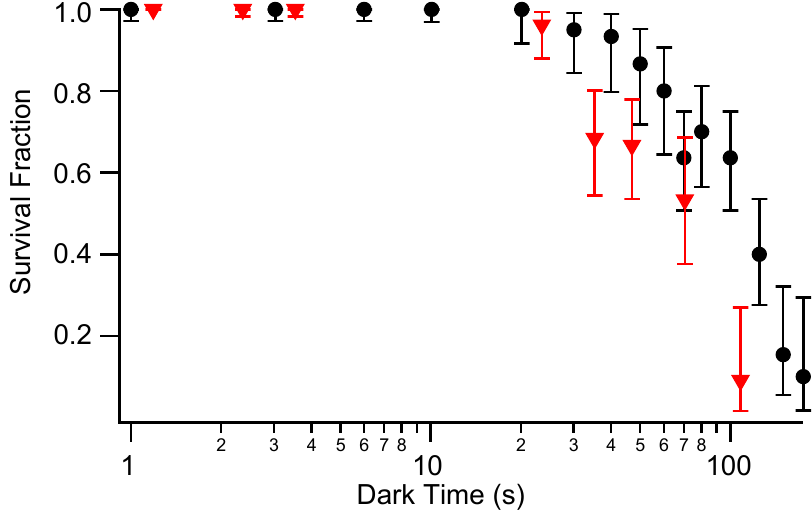}
\caption{\label{fig:DarkLifetime} Survival fraction for stationary ions (black circles) and ions transported (red triangles) at 1 m/s during the experiment dark time. The trap RF is 300 Vpp at 53.17 MHz with axial well depth $\approx 100$ meV.}
\end{figure}

Ion dark lifetimes (survival rates without Doppler cooling) are measured over the non-slotted region of the trap. An acousto-optic modulator (AOM) is normally used for switching the 397 nm cooling beam on and off. To ensure that leakage through the AOM does not affect the measured lifetimes, a mechanical shutter is used during the dark time of this measurement. As shown in Fig. \ref{fig:DarkLifetime}, the ion dark lifetime (50\% survival fraction) is around 100 s.

In previous surface traps \cite{Wright2013}, ions were transported at a speed of 1 m/s using standard air-side DACs with 500 kHz update rate. By contrast, even after optimizing packet sizes as described in Sec. \ref{sec:software}, update rates for the in-vacuum DACs are limited to 40 kHz.  A potential concern with the slower update rate is distortion of the transport waveforms, which must now contain fewer voltage update steps to achieve 1 m/s transport.

Nevertheless, transport tests using the in-vacuum DACs demonstrate effective ion transport for many meters (multiple round trips over a 1 mm region of the trap) without significant loss. In particular, when the ion is transported at 1 m/s for the duration of the dark time, the measured dark lifetime is 70 s (Fig. \ref{fig:DarkLifetime}), corresponding to 70 meters of transport in the dark.   Fluorescence measurements confirm that the ion is indeed transporting out of range of the cooling beam during each round-trip.

\subsection{\label{sec:modes}Mode frequencies}

\begin{figure}[htbp]
\centering
\includegraphics{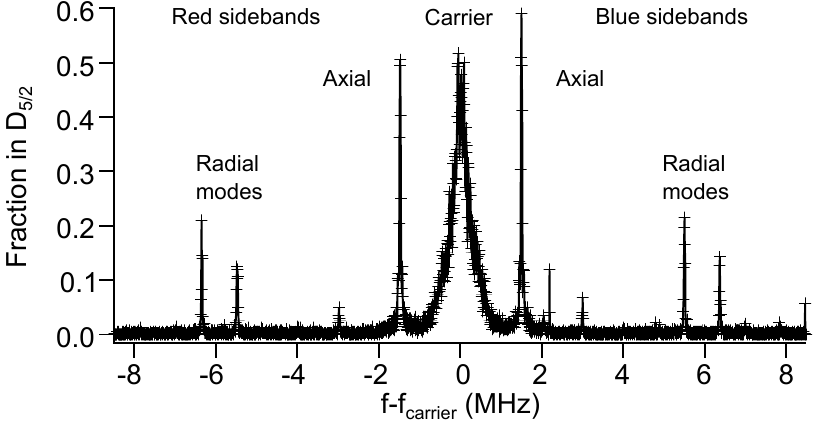}
\caption{\label{fig:OneIonModes}Carrier transition and adjacent motional sidebands for one trapped $^{40}\text{Ca}^+$ ion as a function of the 729 nm laser frequency.  The axial mode frequency is 1.5 MHz with trap RF 300 Vpp at 53.17 MHz (Mathieu parameter \cite{Ghosh1995} $q \approx 0.3$).}
\end{figure}

The $^{40}\text{Ca}^+$ electric quadrupole transition at 729 nm (${^2S_{1/2}},m_s=-1/2 \rightarrow {^2D_{5/2}},m_s=-5/2$) is used to measure ion motional sidebands around a carrier (motion independent) transition. A weak magnetic field of $\approx 8$ gauss lifts the degeneracy of Zeeman sublevels. Figure \ref{fig:OneIonModes} shows the carrier transition with resolved axial and radial sidebands along with cross terms and higher order sidebands. The measured ion axial frequency is 1.5 MHz, with radial secular modes at 5.5 MHz and 6.4 MHz.

\subsection{\label{sec:stability}Secular mode stability}

\begin{figure}[htbp]
\centering
\includegraphics{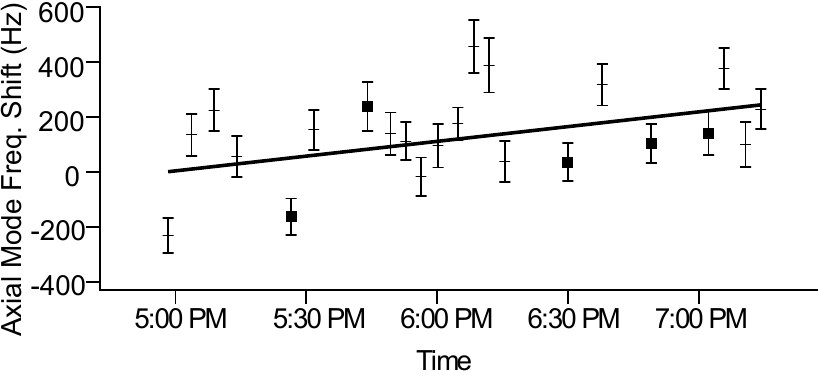}
\caption{\label{fig:ModeStability}Axial mode frequency stability. Measurements just after ion reloads are indicated by solid squares. The linear fit gives a 100(50) Hz/hr drift.}
\end{figure}

Mode frequency stability is a critical factor for gate fidelity in quantum information processing.  The ion axial frequency is measured over the unslotted region of the trap ($z=-390$ $\mu$m, 660 $\mu$m from the load zone center), repeatedly over a two hour period which includes several ion reloads. The resulting set of measurements is shown in Fig. \ref{fig:ModeStability}. The frequency drift is roughly 100(50) Hz/hr, and ion loading does not noticeably shift the mode frequency, as would result from laser charging of the trap surface or drifts in DC trapping potentials. Radial mode stabilities are not measured, as these are set mainly by the RF stability, which is independent of the in-vacuum DAC system.

\subsection{\label{sec:heating}Heating rate}
\begin{figure}[htbp]
\centering
\includegraphics{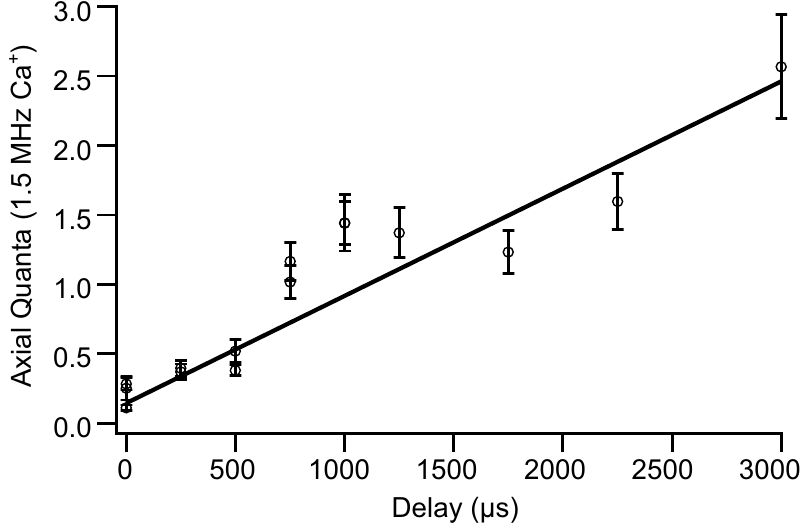}
\caption{\label{fig:HeatingRate} Heating rate measurement showing the number of quanta in the axial mode as a function of time in the dark. The resulting heating rate is 0.8(1) quanta/ms. The ion height is 60 $\mu$m above the trap surface.}
\end{figure}

The heating rate of the ion axial mode (1.5 MHz) is measured over the unslotted region at $z=-390$ $\mu$m. The ion is sideband cooled to an average phonon occupation number of $\bar{n} \approx 0.2$ in the axial mode.  It is then allowed to sit in the dark for a controlled delay time between 0 and 3 ms. Following the delay, the red and blue sidebands are measured and $\bar{n}$ is calculated from $\bar{n}=x/(1-x)$ where  $x=I_{\text{red}}/I_{\text{blue}}$ is the ratio of sideband strengths \cite{Turchette2000}. The results are shown in Fig. \ref{fig:HeatingRate}. A linear fit to the data gives a heating rate of 0.8(1) quanta/ms.

\subsection{\label{sec:Ezmap}Axial field map}

\begin{figure}[htbp]
\centering
\includegraphics{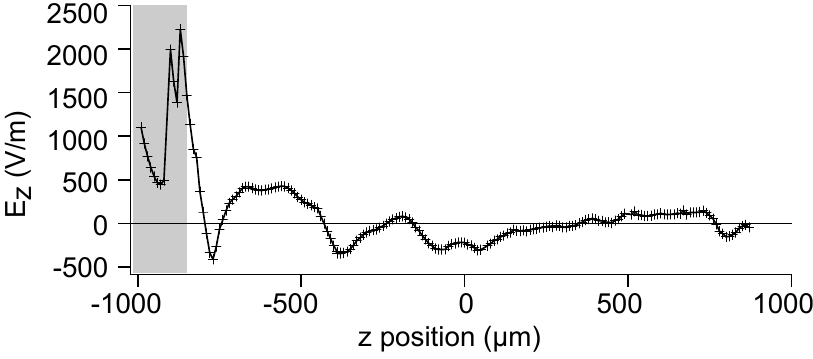}
\caption{\label{fig:AxialScan} Stray axial field $E_z$ measurements over the trap structure. Gray box indicates positions within the loading slot (edge $-850$ $\mu$m). Field magnitudes over the unslotted region are 500 V$/$m or smaller.}
\end{figure}

To test connectedness of all electrodes to their corresponding DAC channels and to probe for possible stray electric fields introduced by the in-vacuum DACs, the ion position and stray axial field strength $E_z$ are mapped out over the length of the trap structure.  Nonzero $E_z$ is indicated by a shift in the ion's axial position as the harmonic trapping potential is scaled in strength.  Results are shown in Fig. \ref{fig:AxialScan}. The ion position is measured via fluorescence imaging, with CCD camera pixel size calibrated to trap features of known length. Electrode failure would result in strong excursions of the ion from the calculated positions; no such excursions are observed.

\subsection{\label{sec:multiples}Multiple ion trapping}

\begin{figure}[htbp]
\centering
\includegraphics{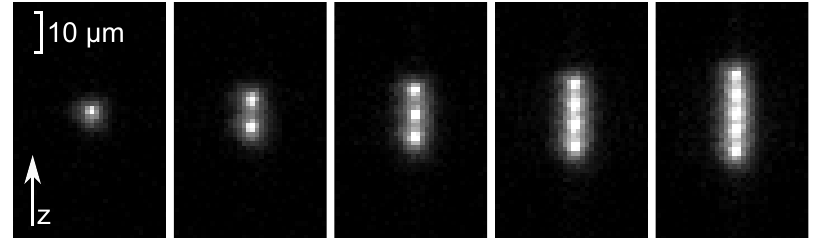}
\caption{\label{fig:MultipleIons}Fluorescence images in the load slot of one to five trapped ions.}
\end{figure}

Quantum information processing requires storage and manipulation of multiple trapped ions. Basic multi-ion capability is demonstrated with the in-vacuum DAC system: loading of up to five ions in a single trapping well, and co-transporting of two ions to the non-slotted region of the trap. Figure \ref{fig:MultipleIons} shows CCD camera images of multiple trapped ions above the load slot. We regularly transport two ions out of the load zone and resolve motional sidebands corresponding to center-of-mass and stretch axial modes (frequencies $\omega_z$ and $\sqrt{3}\omega_z$, respectively).

\section{Summary and Conclusions}

In-vacuum electronics reduce the number of electrical feedthroughs required to control a 78-electrode microfabricated ion trap by nearly an order of magnitude.  The design scales favorably to more complex trap geometries, with each additional 40-channel DAC requiring only one additional pair of vacuum feedthrough lines.  Commercially available integrated circuits are decapsulated from original packaging to produce fully UHV-compatible trap and regulator circuit boards.  A serial interface allows communication with an air-side computer and controller board.

The in-vacuum electronics are used successfully to control loading and manipulation of $^{40}\text{Ca}^+$ ions in a GTRI Gen V surface-electrode ion trap.  Trap performance is characterized by axial mode stability, heating rate, stray axial electric fields, and ion dark lifetime.  The integrated system performs comparably to earlier traps with standard air-side electronics \cite{Doret2012}, demonstrating the potential of this approach to simplify hardware requirements for the increasingly complex schemes of trapped-ion quantum computing.

The large die area of the commercial DACs with respect to the active region of the ion trap will place an eventual limit on the number of channels that can be driven in a single electronics package.  Increasing trap complexity would require reducing the DAC die area and developing a higher connection-density alternative to wirebonds for the trap chip. Scaling could also be improved by incorporating multiplexers in the in-vacuum circuitry. Higher-voltage operation in the 10 to 100 V range could be achieved by following the in-vacuum DACs with decapsulated high-voltage amplifiers supplied by an additional pair of power feedthroughs. However, the die footprint and power dissipation of the high voltage amplifiers might limit the number of channels that could be amplified. Integration with cryogenic ion trapping systems \cite{Vittorini2013,Antohi2009} would pose the additional challenge of thermally isolating the in-vacuum electronics from the trap structure and maintaining the components at high enough temperatures to operate.

\begin{acknowledgments}
This material is based upon work supported by the Office of the Director of National Intelligence (ODNI), Intelligence Advanced Research Projects Activity (IARPA) under The Space and Naval Warfare Systems Command (SPAWAR) contract number N6600112C2007. All statements of fact, opinion, or conclusions contained herein are those of the authors and should not be construed as representing the official views or policies of IARPA, the ODNI, or the U.S. Government.
\end{acknowledgments}

\appendix


\providecommand{\noopsort}[1]{}\providecommand{\singleletter}[1]{#1}%

\end{document}